\documentclass[12pt,preprint]{aastex}
\usepackage{apjfonts,graphicx,ifpdf}

\def\etal{{\it et~al.}}
\newcommand{\mum}{$\mu$m}
\newcommand{\HII}{H\,{\sc ii}}

\newcommand{\mjb}{mJy~beam$^{-1}$}

\newcommand{\kms}{km~s$^{-1}$}
\def\hto{{\hbox{H$_{2}$O}}}

\shorttitle{Water Masers in the Antennae Galaxies}
\shortauthors{Brogan, Johnson, \& Darling}

\begin{document}

\title{Water Masers Associated with Star Formation in the Antennae Galaxies}

\author{Crystal Brogan\altaffilmark{1}, \&
Kelsey Johnson\altaffilmark{2,3}, Jeremy Darling\altaffilmark{4}}

\altaffiltext{1}{National Radio Astronomy Observatory, 520 Edgemont
  Rd, Charlottesville, VA 22903; cbrogan@nrao.edu}
\altaffiltext{2}{Department of Astronomy, University of Virginia,
PO Box 3818, Charlottesville, VA  22903}
\altaffiltext{3}{KEJ is also an Adjunct Astronomer at the National Radio 
Astronomy Observatory}
\altaffiltext{4}{Center for Astrophysics and Space Astronomy,
Department of Astrophysical and Planetary Sciences,
University of Colorado, 389 UCB, Boulder, CO 80309-0389}

\begin{abstract}

  We present Very Large Array observations with 80 milliarcsecond
  resolution ($\sim 9$ pc) of the recently discovered Galactic-analog
  \hto\/ masers in the Antennae interacting galaxies
  (NGC~4038/NGC~4039; Arp244). Three regions of water maser emission
  are detected: two in the ``interaction region'' (IAR) and the third
  $\sim 5.6''$ ($\gtrsim 600$~pc) west of the NGC~4039 nucleus. The
  isotropic \hto\/ maser luminosities range from 1.3 to 7.7
  $L_{\sun}$. All three maser regions are mostly obscured in the
  optical/near-infrared continuum, and are coincident with massive
  CO-identified molecular clouds. The \hto\/ maser velocities are in
  excellent agreement with those of the molecular gas. We also present
  archival VLA 3.6~cm data with $\sim 0\farcs28$ ($\sim 30$~pc) and
  $\sim 0\farcs8$ ($\sim 90$~pc) resolution toward the maser
  locations. All three maser regions are coincident with compact
  3.6~cm radio continuum emission, and two are dominated by thermal
  ionized gas, suggesting the presence of natal super star clusters
  containing the equivalent of a few thousand O stars. We also present
  detailed comparisons between the radio data and existing HST ACS
  (optical) and NICMOS (near-IR) data and find that both maser regions
  in the IAR are also associated with Pa$\alpha$ emission and neither
  source is detected shortward of 2 \mum\/. These results highlight
  the potential of using Galactic-analog \hto\/ masers to pinpoint
  sites of young super star cluster formation with exquisite angular
  resolution.

\end{abstract}


\section{Introduction}

Water masers are found in the vicinity of $\sim 70\%$ of infrared
bright (100 \mum\/ $>$ 1000 Jy and 60 \mum\/ $>$ 100 Jy) ultracompact
\HII\/ (UCHII) regions in our Galaxy \citep{churchwell90,kurtz05},
providing excellent signposts for active star formation.  The term
``kilomaser'' was coined to describe extragalactic H$_2$O masers with
luminosities comparable to the brightest star formation H$_2$O masers
in our Galaxy (e.g.  W49N; $L_{\rm H_2O}\sim 1$ L$_\odot$). Kilomaser
isotropic luminosities ($L_{\rm H_2O} <10$ L$_\odot$) are much lower
than the more widely studied ``megamasers'' found in the nuclear
regions of AGN (i.e.  NGC~4258), with luminosities up to 10$^4$
L$_\odot$ \citep{barvainis05}. Since ``kilomasers'' can in principle
be either of nuclear (i.e. amplifying a background AGN) or star
formation origin, we will use the term ``Galactic-analog'' (GA) \hto\/
maser to refer to non-nuclear masers with kilomaser luminosity. Natal
super star clusters are thought to contain 1000s of UCHII regions,
therefore GA-\hto\/ masers may help pinpoint sites of young
extragalactic cluster formation, though there have been few searches
with the required sensitivity. Until recently GA-\hto\/ masers had
been unambiguously detected toward only six nearby galaxies: LMC, M82,
IC~342, IC~10, M33, and NGC~2146 with isotropic luminosities in the
range 0.005 to 4 $L_{\sun}$ \citep[][and references
therein]{whiteoak86,henkel05,castangia08}. The remaining few known
\hto\/ kilomasers are either associated with AGN activity or their
origin is ambiguous.

We recently conducted a Greenbank Telescope\footnote{The National
  Radio Astronomy Observatory operates the GBT and VLA and is a
  facility of the National Science Foundation operated under
  cooperative agreement by Associated Universities, Inc.}  (GBT)
search for \hto\/ masers toward four nearby starburst galaxies ($3< D<
22$ Mpc), known to harbor natal super star clusters and without known
AGN (or at worst low luminosity AGN) down to a sensitivity level
sufficient to detect strong GA-\hto\/ masers. Positive detections were
found for all four galaxies \citep[the Antennae galaxies, He2-10,
NGC~4214, and NGC~5253;][]{darling2008}, suggesting that GA-\hto\/
maser emission may be common in starburst galaxies. The unusual
success of this mini-survey arises from its order of magnitude greater
sensitivity compared to most previous single dish surveys
for either kilomasers or megamasers. 

The strongest masers were detected toward the Antennae (D$\sim22$
Mpc\footnote{Antennae distance estimates range from $13.8\pm 1.7$ Mpc
\citep{saviane08} to $22\pm 3$ Mpc \citep{schweizer08}; we adopt 22
Mpc.})  interacting galaxies (NGC~4038/NGC~4039; Arp244). Indeed, this
violent merger is an ideal laboratory to study how \hto\/ maser
emission behaves in regions of extreme star formation.  In this paper
we report sensitive high angular resolution Very Large Array (VLA)
water maser observations of the Antennae galaxies, along with archival
3.6 and 6~cm continuum, CO, and optical/near-IR data in order to
pinpoint the maser locations and determine the nature of the emission.

\section{VLA Observations and Data Reduction}\label{sec:obs}

We observed the ($6_{16}-5_{23}$) ortho-water maser line at
22.23508~GHz toward the Antennae galaxies in the VLA A-configuration
using fast-switching and reference pointing.  The full width at half
power (FWHP) at the Doppler shifted frequency of $\sim 22.11$ GHz is
$\sim 2\arcmin$, encompassing both nuclei (NGC~4038 and NGC~4039), as
well as the interaction region (IAR; see Figure~\ref{spectra}a); in
comparison the GBT primary beam was only $30\arcsec$. In order to
cover the velocity extent of maser emission detected with the GBT with
similar velocity resolution (3.7 \kms\/) required three correlator
settings, each with 2 IFs. We also retrieved A and B-configuration
3.6~cm continuum data, as well as BnA-configuration 6~cm data from the
VLA archive \citep[the B-configuration data were included in the
  multi-configuration study of][]{neff00}. Further details of the VLA
data are given in Table~\ref{obs}. Data calibration followed standard
high-frequency procedures in AIPS, including using a model for the
brightness distribution of the absolute flux calibrator 3C~286;
absolute flux calibration is good to $\sim 5\%$. The line-free
channels from the $\sim 22.11$ GHz data were used to estimate the
1.3~cm continuum emission in the UV-plane. The maser line data were
subsequently imaged in CASA (to facilitate regridding of the six IFs),
and primary beam correction was applied. The continuum data were
imaged in AIPS, and primary beam corrected. The astrometric accuracy
is better than $\sim 0\farcs05$, while the {\em relative} position
uncertainty of the maser data is an order of magnitude better). All
velocities are presented in the barycenter frame, optical definition.

\begin{deluxetable}{ll}
\tabletypesize{\scriptsize}
\tablewidth{0pc}
\tablecaption{VLA Observing Parameters\label{obs}}
\tablecolumns{2}
\tablehead{
\colhead{Parameter}  & \colhead{Value} }
\startdata
\cutinhead{\hto\/ maser and 1.3~cm (22.11 GHz) continuum Observations}
Project (config.) & AB1304 (A) \\
Observing Dates & 10, 11, 15 Nov 2008  \\
Bandwidth & 6 $\times$ 6.25 MHz \\
%Central frequencies & 22.1104, 22.1150, 22.1197, 22.1244, 22.1292, 22.1338 %GHz \\
%Central velocities & 1700, 1636, 1573, 1509, 1446, 1382 \kms\/ \\
Velocity Bandwidth & 1360 to 1740 \kms\/ \\
Velocity resolution & 3.7 \kms\/ \\
Spec. line resolution & 108 mas $\times$ 61 mas
(P.A.$=+1.8\arcdeg$) \\
Spec. line rms noise & 1 \mjb\/ \\
Cont. resolution$^{a}$ & 350 mas $\times$ 220 mas (P.A.$=10\arcdeg$); 
$\lesssim 1000$ k$\lambda$\\
Cont. rms noise & 0.14 \mjb\/ \\
\cutinhead{Archival 3.6~cm (8.46 GHz) Continuum Observations}
Project (config.) & AP478 (A), AS796 (A), AA301 (A), AN079 (B) \\
Observing Dates &  01 Nov, 21 Dec 2004; 09 Mar 2006; 07 Sep 1998 \\
Bandwidth & $2 \times$ 50 MHz \\
%Central frequencies  & 8.4351, 8.4851 GHz \\
Cont. resolution (A) & 350 mas$\times 220$ mas (P.A.$=10\arcdeg$); $>30$
k$\lambda$ \\
Cont. rms noise (A) & 0.032 \mjb\/ \\
Cont. resolution (B) & $1\farcs03 \times 0\farcs67$( P.A.$=-2.4\arcdeg$) \\
Cont. rms noise (B) & 0.013 \mjb\/ \\
\cutinhead{Archival 6~cm (4.85 GHz) Continuum Observations}
Project (config.) & AN074 (BnA) \\
Observing date & 31 Jan 1997 \\
Bandwidth & 2 $\times$ 25 MHz \\
%Central frequencies & 4.8726, 4.8226 GHz \\
Cont. resolution& $1\farcs03 \times 0\farcs67$ (P.A.$=82.1\arcdeg$); $>65$
k$\lambda$ \\   
Cont. rms noise & 0.022 \mjb\/ 
\enddata\tablecomments{A Briggs weighting of robust=0 was used unless
  otherwise specified. UV weighting adjustments are indicated after
  the resolution; $\lesssim$ indicates a Gaussian taper and $>$ indicates a cut.}
\tablenotetext{a} {Natural weighting was used.}
\end{deluxetable}

\section{Results}\label{sec:results}

\subsection{\hto\/ Maser Emission}

\begin{figure}
\epsscale{0.7}
\plotone{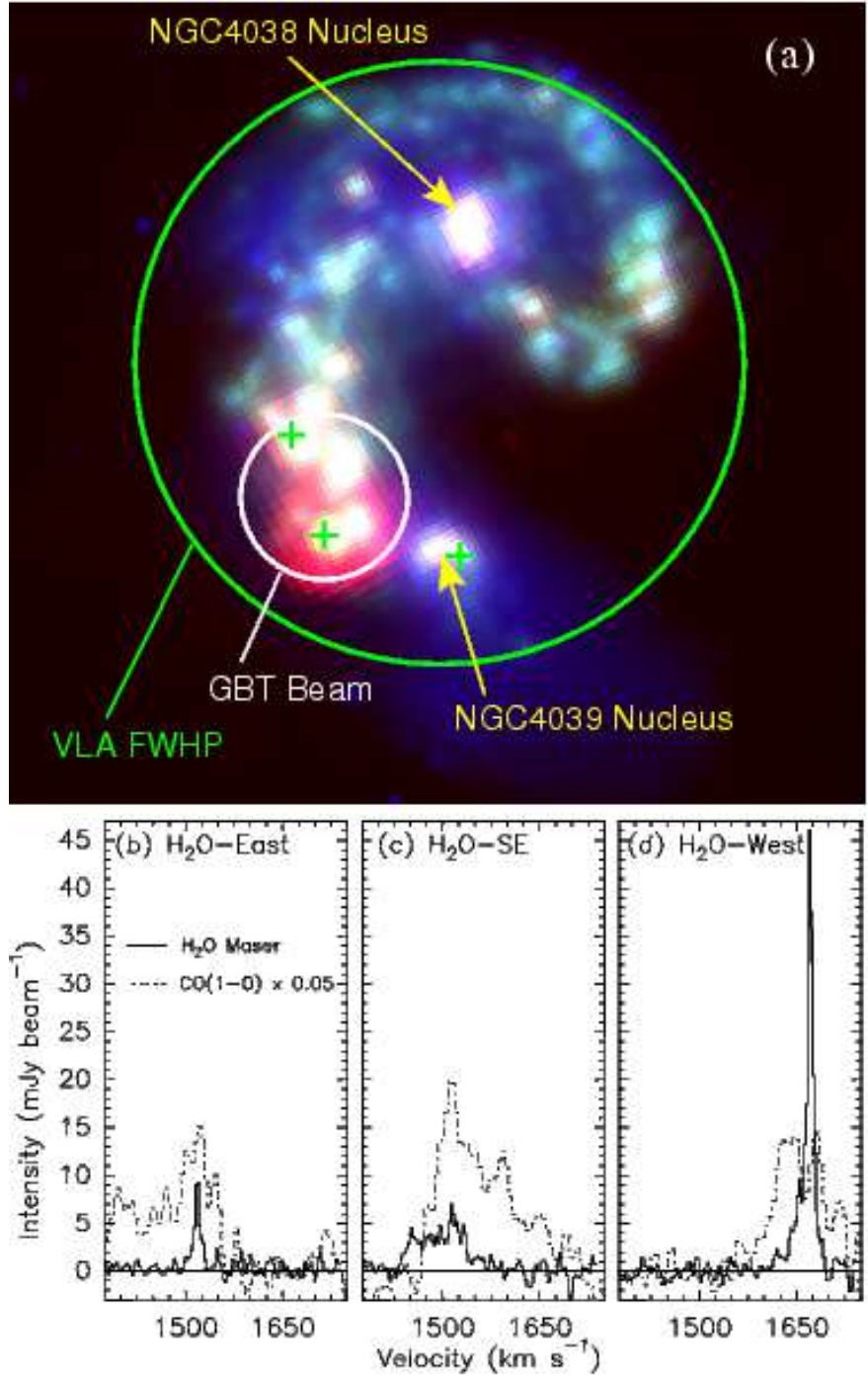}
\caption{(a) Three color {\em Spitzer} image of the Antennae galaxies
  with RGB mapped to 24, 8, \& 3.6 \mum\/. Green $+$ symbols show the
  locations of the three \hto\/ maser regions detected by the VLA. In
  panels (b), (c), and (d) the VLA \hto\/ maser spectra are shown by
  the solid histogram, and $\sim 4\arcsec$ resolution CO(1--0) spectra
  from \citet{wilson00} are superposed as dot-dash lines. 
\label{spectra}}
\end{figure}

As shown on the three-color {\em Spitzer} composite in
Figure~\ref{spectra}a, three distinct regions of \hto\/ maser emission
are resolved by the VLA; two are located in the 24~\mum\/-bright IAR
between the two nuclei, and one is located $5\farcs6$ ($\gtrsim
600$~pc) to the west of the NGC~4039 nucleus. For ease of referencing,
these three maser regions are denoted \hto\/-East, \hto\/-SE, and
\hto\/-West. We call these ``maser regions'' because at the current
resolution $\sim 80$ mas ($\sim 9$ pc), the observed emission is most
likely the sum of many individual maser spots. Spectra of the maser
emission are shown in Figures~\ref{spectra}b,c, and d, and the
observed properties are given in Table~\ref{lines}. At the current
angular resolution, no variation of maser position with velocity is
detected.  The range of maser velocities detected by the VLA is in
excellent agreement with that of the GBT data, despite the 16$\times$
larger area covered in the VLA data. Differences in intensity between
the VLA and GBT data can be explained by the maser offsets from the
GBT pointing center (see Fig.~\ref{spectra}a).  Using the observed
line properties (Table~\ref{lines}) we find isotropic \hto\/ maser
luminosities ranging from 1.3 to 7.7 $L_\odot$, which is on the high
side, but not dissimilar to the other galaxies with known GA-\hto\/
emission (see \S 1). The \hto\/ luminosity sensitivity limit is $\sim
0.6$ $L_\odot$ (assumes $V_{peak}$=5 mJy ($5\sigma$) and $\Delta
V_{FWHM}$=10 \kms\/), brighter than the majority of Galactic water
masers.

\begin{deluxetable}{lccccccl}
\tabletypesize{\scriptsize}
\tablecaption{VLA Water Maser Properties\label{lines}}
\tablewidth{0pt} \tablehead{
\colhead{Maser} & 
\multicolumn{2}{c}{Position} &
\colhead{$V_{range}$} & 
\colhead{$V_{peak}$} &
\colhead{$S_{peak}$} & 
\colhead{$\int S{\rm d}v$} &
\colhead{$L_{H_2O}$}\\
\colhead{Region} & 
\colhead{$\alpha$ (J2000)} &
\colhead{$\delta$ (J2000)} &
\colhead{(km s$^{-1}$)} & 
\colhead{(km s$^{-1}$)} & 
\colhead{(mJy)} & 
\colhead{(mJy km s$^{-1}$)} & 
\colhead{($L_\odot$)} }
\startdata
\hto\/-East  & 12:01:55.4590 (2) & $-$18:52:45.653 (5)   &  
              1502 - 1528 & 1520.4 & 9.3 (0.9)  &  116 (12) & 1.3
              (0.1) \\ 
% channels 82 - 92
\hto\/-SE    & 12:01:54.9959 (4) & $-$18:53:05.543 (8)   &
              1439 - 1539 & 1515.1 &  7.2 (1.0)   & 370 (25) & 4.1 (0.4) \\
% channels 78 - 116
\hto\/-West  & 12:01:53.1257 (1) & $-$18:53:09.805 (2)   &
              1639 - 1689 & 1670.5 & 44.3 (0.1) & 690 (25) & 7.7 (0.5) 
% channels 21 - 40     
\enddata \tablecomments{Positions and peak flux densities measured
  from 2-D Gaussian fits. Isotropic line luminosities were computed
  from $L_{H_2O} = (2.3\times 10^{-5}\ L_\odot)\times D^2\times\int
  S{\rm d}v$, where the distance $D$ was assumed to be 22 Mpc and
  $\int S{\rm d}v$ is in mJy \kms\/ \citep{henkel05}.  }
\end{deluxetable}

% V_LSR = V_HEL - 3.4

\subsection{Radio Continuum Emission
\label{Radio_Text}}

\begin{figure*}
\epsscale{1.0}
\plotone{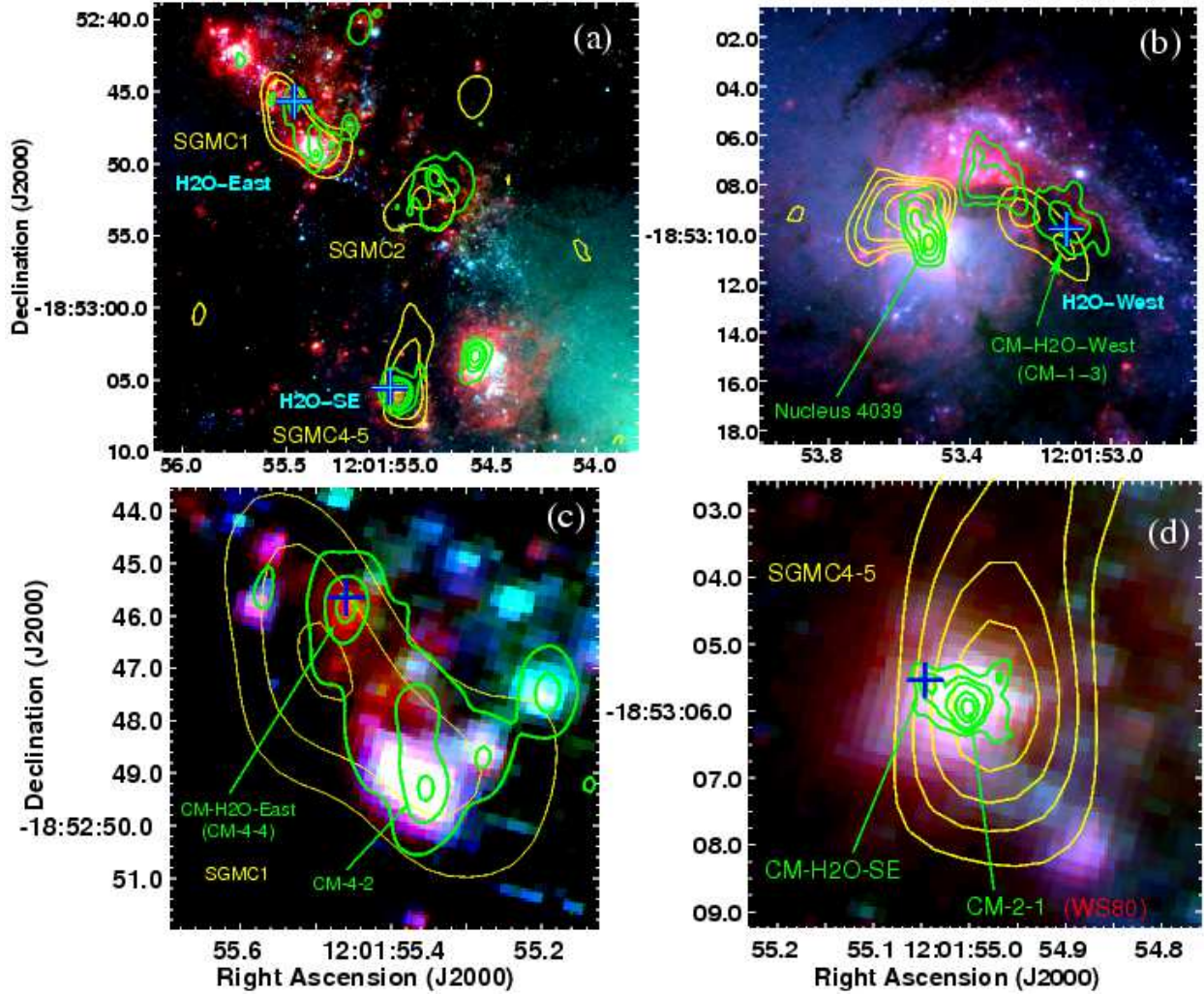}
\caption{Three color images of (a) the IAR, (b) NGC 4039 nucleus, (c)
  \hto\/-East, and (d) \hto\/-SE. For (a) and (b) RGB is mapped to HST
  ACS optical filters H$\alpha$, 814W, \& 550M. For (c) and (d) RGB is mapped
  to HST NICMOS near-IR filters Pa$\alpha$, F160, \& F110. In all four panels
  blue $+$ symbols show the \hto\/ maser locations and the yellow
  contours show SMA CO(3--2) integrated intensity
  \citep{petitpas07}. Green contours show VLA 3.6 cm $0\farcs8$ resolution
  continuum emission in (a), (b), and (c) with contour levels of 0.05,
  0.1, 0.3, 0.6, and 1.2 \mjb\/. Green contours show VLA 3.6 cm $0\farcs28$ resolution  continuum emission in (d) with contour levels of 0.1, 0.2, 0.3, 0.6, and 0.9 \mjb\/. The SGMCs identified by \citet{wilson00} and the
  cm-$\lambda$ continuum sources identified by \citet{neff00} are
  indicated.
  \label{HST}}
\end{figure*}

We have created matched resolution $\sim 0\farcs8$ ($\sim 90$ pc) 3.6
and 6~cm continuum images with high signal-to-noise, as well as higher
resolution $\sim 0\farcs28$ ($\sim 30$ pc) 1.3 and 3.6~cm images,
albeit with significantly less sensitivity. Care was used to match the
UV-coverage of the continuum pairs at both resolutions. These images are
the highest resolution radio images available to date towards the
Antennae \citep[see for example the $\sim 1.1\arcsec$ resolution
  3.6~cm VLA data of][]{neff00}. A complete analysis of these data
throughout the Antennae will be presented in a forthcoming paper, for
now we concentrate on the maser locations.  Radio continuum sources
associated with \hto\/-East and \hto\/-West are detected in the $\sim
0\farcs8$ resolution 3.6 and 6~cm images \citep[also see][their
  sources 4-4 and 1-3 respectively]{neff00}. We have also resolved
compact continuum emission towards \hto\/-SE in the $\sim
0\farcs28$ 1.3 and 3.6~cm images \citep[this source is not resolved
  from the bright][source 2-1 at poorer resolutions]{neff00}. The
radio source associated with \hto\/-East is also detected in the
higher resolution (poorer sensitivity) 1.3 and 3.6~cm continuum
images, though \hto\/-West is not.  The 3.6~cm continuum contours are
shown in Figures~\ref{HST}a,b,c,d (see \S4.1).

The radio continuum sources are denoted CM-\hto\/-East, CM-\hto\/-SE,
and CM-\hto\/-West, and their observed flux densities and spectral
indices ($S_{\nu}\propto \nu^{\alpha}$) are listed in
Table~\ref{RadioTab}. The spectral index for CM-\hto\/-East is flat
and in good agreement with \citet[][their source 4-4,
$\alpha=-0.16\pm 0.19$]{neff00}.  We also find a flat (or possibly
inverted) spectrum for CM-\hto\/-SE (detected here for the first
time).  In contrast, we find that CM-\hto\/-West has a steep spectrum
($\alpha=-0.6^{+0.4}_{-0.4}$), while \citet{neff00} inferred an
inverted spectrum (their source 1-3, $\alpha=+0.38\pm 0.19$). This
inconsistency can be explained by the fact that these authors use a
fitted size for this source at 3.6~cm, but only a peak flux density at
6~cm due to a poor Gaussian fit, thus underestimating the 6~cm flux
density. Since both CM-\hto\/-East and CM-\hto\/-SE have flat radio
spectra, and are coincident with Pa$\alpha$ emission (see\S4.1), both
are likely to be dominated by thermal \HII\/ regions. In contrast,
CM-\hto\/-West has a steep spectrum and thus appears to be dominated
by non-thermal emission; broader radio wavelength coverage is required
to disentangle a thermal component for this maser region.

For the two thermal sources, CM-\hto\/-East and CM-\hto\/-SE, the
production rate of ionizing photons ($Q_{Lyc}$) can be estimated using
Equation 2 from \citet{condon92} assuming thermal, optically thin
emission \citep[see][]{johnson09}.  Although the 1.3~cm data would
generally be preferred (to exclude non-thermal contamination, and to
have a higher likelihood of being optically thin), we have used the
3.6~cm data to determine Q$_{LyC}$ due to its superior sensitivity
(see Table~\ref{obs}).  An O7.5V star produces an ionizing flux of
Q$_{LyC}$ =$10^{49}$~s$^{-1}$ \citep[hereafter
  O*;][]{leitherer,vacca96}, suggesting that the ionized gas
associated with the two IAR maser regions is equivalent to $\sim 2000
- 5000$ O* stars.  Note that these Q$_{LyC}$ values could be
underestimates if a significant fraction of the ionizing flux is
absorbed by dust or able to escape from the \HII\ region
\citep[see][]{johnson03,reines08,johnson09}.

%\begin{equation}
%{{Q_{Lyc}}}
%\geq6.3\times10^{52}~{{\rm s}^{-1}}
%\Big({{T_e}\over{10^4~{\rm K}}}\Big)^{-0.495}
%\Big({{\nu}\over{{\rm Ghz}}}\Big)^{0.1}
%\Big({{L_{thermal}}\over{10^{27}~{\rm erg~s^{-1}~Hz}^{-1}}}\Big).
%\end{equation}

\begin{deluxetable}{lccccc}
\tablecaption{Radio Continuum Properties \label{RadioTab}}

\tablehead{\colhead{Source} & 
\colhead{F$_{\rm 6.0 cm}$\tablenotemark{a}} & 
\colhead{F$_{\rm 3.6_cm}$\tablenotemark{a}} &
\colhead{F$_{\rm 1.3 cm}$\tablenotemark{a}} & \colhead{$\alpha$\tablenotemark{b}} & \colhead{Q$_{LyC}$\tablenotemark{c}} \\
\colhead{} & \colhead{(mJy)} & \colhead{(mJy)} & \colhead{(mJy)} & \colhead{} & \colhead{($\times 10^{52}$ s$^{-1}$)} }
\startdata
CM-\hto\/-East\tablenotemark{d} & 1.1 (0.1)  & 1.00 (0.08) & \nodata   & $-0.17^{+0.2}_{-0.2}$ & 5.1 \\
CM-\hto\/-SE\tablenotemark{e}   & \nodata      & 0.5 (0.1)   & 0.6 (0.4) & $+0.2^{+0.9}_{-0.4}$ & 1.9 \\
CM-\hto\/-West\tablenotemark{d} & 0.9 (0.14)  & 0.64 (0.09) & \nodata   & $-0.6^{+0.4}_{-0.4}$ & ?
\enddata
\tablenotetext{a}{Flux densities measured from polygon regions;
  uncertainties are $({\rm no.~independent~beams})^{0.5}\times 3\sigma$ plus a $5\%$ calibration uncertainty
  (added in quadrature); blanks mean source was resolved out or
  too confused.}
\tablenotetext{b}{Uncertainties are 68$\%$ confidence levels
  (1$\sigma$); for  \hto\/-SE 0.4 mJy is
  a lower limit for the 1.3~cm flux density, so this defines the lower
  bound on its $\alpha$.}
\tablenotetext{c}{Inferred from 3.6~cm flux densities.}
\tablenotetext{d}{Flux densities measured from the $\sim 0.8\arcsec$ data.}
\tablenotetext{e}{Flux densities measured from the $\sim 0.28\arcsec$ data.}
\end{deluxetable}

% CO contour levels on east 60, 95, 130, 165, 200
% CO contour levels on west 60, 75, 90, 120, 150

\section{Discussion}\label{sec:discussion}

\subsection{Multiwavelength Comparison of the Maser Environments}

To further assess the nature of the sources associated with the maser
emission, we have obtained existing {\em Hubble Space Telescope} (HST)
ACS and NICMOS data, along with Owens Valley Radio Observatory (OVRO)
CO(1--0), and Submillimeter Array (SMA) CO(3--2) molecular line data
\citep{whitmore99,wilson00,petitpas07}. To use these data we must
first ensure the registration of the different data sets. The
astrometry of the VLA data is good to $0\farcs05$, far superior to the
other available data. Because of their shorter wavelengths, the OVRO
and SMA CO positions are likely less accurate ($\sim 0\farcs2$). The
HST data have an astrometric uncertainty of $\sim 1\arcsec$. For
example, by comparing with the \citet{neff00} radio data,
\citet{whitmore_zhang02} found a $1\farcs2$ offset for their WFPC2
images. For the current analysis we first matched the
continuum-subtracted Pa$\alpha$ emission in the IAR to the 3.6~cm
continuum since they have the same physical origin.  We then applied
that astrometric solution to the non-continuum-subtracted Pa$\alpha$
image, which has many features in common with the other
optical/near-IR images, allowing us to register and correct all the
other HST images. The resulting shifts for the ACS and NICMOS data,
are $-0\farcs97$, $+1\farcs58$; and $+0\farcs49$, $+0\farcs20$,
respectively \citep[we also find good agreement with the WFPC2 offset
determined by][]{whitmore_zhang02}. We estimate that the final
relative astrometry is good to $\sim 0\farcs2$.

Figures~\ref{HST}a, and b show three-color optical wavelength HST
images of the IAR, and NGC~4039 nuclear region, respectively, and are
superposed with the locations of the masers, 3.6~cm contours, and SMA
CO(3--2) integrated intensity contours from
\citet{petitpas07}. Figures~\ref{HST}c, and d further zoom into the
IAR to show close-up near-IR NICMOS views of \hto\/-East and
\hto\/-SE, respectively. Additionally, superposed on the \hto\/
spectra in Figures~\ref{spectra}b,c, and d are scaled OVRO CO(1--0)
spectra from \citet[][these data have significantly higher S/N than
  the CO(3--2) data]{wilson00}. The velocities of 
all three maser regions are encompassed by the velocity range of the CO(1--0)
emission, suggesting that the masers are located within the thermal
molecular gas. Indeed, \hto\/-East and \hto\/-SE are coincident with
super giant molecular complexes (SGMC) 1 and 4-5, respectively
\citep[see][]{wilson00}, and all three maser regions are coincident
with CO(3--2) clouds \citep[also see][]{petitpas07}. Additionally, all
three maser regions are notably dim at optical wavelengths, suggesting
high obscuration (see Fig.~\ref{HST}a, b). Below we discuss the
properties of each region in detail.

\subsubsection{\hto\/-East Region} 

As shown in Fig.~\ref{HST}a and discussed in \S3.2, the \hto\/-East
maser is coincident with a massive thermal \HII\/ region, but is
located in an optically uninteresting region between two optically
bright star forming regions $\sim 4\arcsec$ ($\gtrsim 430$ pc) to the
NE and SW. In the following, we will denote the \HII\/ region itself
as \hto\/-East. \hto\/-East is also coincident with a source of
compact Pa$\alpha$ emission detected by NICMOS (Fig.~\ref{HST}c),
though it is not detected in either of the NICMOS F110W, or F160W
continuum images (it is weakly detected in the F220M image, not
shown). The \hto\/-East cluster is not included in the
\citet{brandl05} WIRC J and Ks band study of the reddest Antennae
clusters because it is so red that it was not detected at J-band.
Notably \hto\/-East is only $\sim 1\farcs5$ west of the 2nd reddest
Antennae cluster \citep[ID$_{\rm WIRC}$=190, $A_v\sim
10$;][]{brandl05}. In contrast, the optically bright massive clusters
located $\sim 4\arcsec$ to the NE and SW of \hto\/-East show
relatively little extinction \citep[$A_v=1.3$ for the SW
sources][]{whitmore_zhang02}.  Unfortunately, the resolution of
existing mid-IR ($\lambda> 2.2$ \mum\/) imaging has been too poor to
resolve the \hto\/-East region from the nearby clusters.

~\hto\/-East has also been detected in Br$\gamma$ by \citet[][ their
  source ``D2'']{gilbert07} using NIRSPEC on KeckII. These authors find
a peak Br$\gamma$ velocity of 1487 \kms\/ and a $\Delta V_{FWHM}$= 55
\kms\/, which is somewhat blue-ward of the \hto\/-East maser and
CO(1--0) peak velocity of 1520 \kms\/ (see Fig.\ref{spectra}b),
suggesting the Br$\gamma$ originates in outflowing ionized gas in
front of the maser region. \citet{gilbert07} find that the bright
cluster to the SW \citep[their source ``D'', and][source
  CM-4-2]{neff00} has a Br$\gamma$ velocity offset by 100 \kms\/ from
\hto\/-East, suggesting that these clusters may be separated
significantly along the line of sight in addition to the $\sim
4\arcsec$ ($\gtrsim 430$ pc) plane-of-sky distance.

%\footnote{Note \citet{gilbert07} misquote the \citet{whitmore99}
%    ``Star 4'' reference position by -0.2s in R.A.}

\subsubsection{\hto\/-SE Region}

The \hto\/-SE region is located only $0\farcs5$ ($\sim 60$ pc) to the
NE of the most massive ($\sim 5 \times 10^6$~M$_\odot$) and brightest
mid-IR to cm-$\lambda$ star cluster ($=$ WS80, B1, CM-2-1, WIRC157,
Peak~1) in the whole of the Antennae galaxies \citep[see for
  example][we denote this cluster as WS80 for ease of
  referral]{whitmore_zhang02,gilbert07,neff00,brandl05,brandl09}. Though
the optical HST data have more than sufficient resolution, \hto\/-SE
is not detected. The majority of longer wavelength data do not have
sufficient resolution, so that only the integrated properties of
\hto\/-SE plus the WS80 cluster are available. Indeed, the \hto\/-SE
region has been resolved and detected in only one other study -- that
by \citet{snijders06} using the VLT VISIR instrument at 12.81 \mum\/
([NEII] + continuum, their source ``1b''). The \hto\/-SE source
contributes about 1/5 of the total flux of the combined \hto\/-SE +
WS80 regions at this wavelength, similar to what is seen in the
$0\farcs28$ resolution 3.6~cm data.

\citet{wilson00} find this region to be coincident with the confluence
of two SGMCs, and as seen in Fig.~\ref{spectra}c, both the maser and
CO(1--0) show particularly broad emission.  \citet{gilbert07} find a
Br$\gamma$ velocity for WS80 (their source ``B1''), of
$v_{peak}=1476$~\kms\/ with a FWHM$=70$~\kms\/.  These values are in
excellent agreement with the \hto\/-SE maser kinematics (1515.1
\kms\/, and 100 \kms\/), suggesting that \hto\/-SE and WS80 are in
close proximity kinematically as well as spatially.
 
Estimated ages for the WS80 cluster range between $1-3.5$ Myr
\citep{whitmore_zhang02,gilbert07,snijders07}, though it is difficult
to pinpoint cluster ages of $\lesssim 3$~Myr due to the lack of
evolution in colors and ionizing fluxes. \citet{whitmore_zhang02}
infer an extinction value of $A_V = 7.6$ for WS80, and
\citet{gilbert07} find $A_K = 1.2$, or roughly $A_V = 10$ for a simple
screen model. Interestingly, in order to account for the lack of
detection of \hto\/-SE at shorter wavelengths, \citet{snijders06}
suggest that $A_v> 72$ for \hto\/-SE!  Given its very high extinction,
\hto\/-SE may be an extremely young super star cluster in the making,
though it is difficult to distinguish between extreme youth (and thus
small physical size and $Q_{LyC}$ compared to WS80) vs. a simply less
massive (and thus less able to clear its natal material) but more
evolved cluster.

\subsection{\hto\/-West Region}

The western maser resides in an optically thick dust lane coincident
with a CO(3--2) molecular cloud, and is virtually invisible at every
other available wavelength except for weak radio emission (see
Fig.~\ref{HST}b).  Moreover, because the radio emission is dominated
by non-thermal emission it is impossible to discern the nature of any
ionized thermal emission that might be associated with the maser
region. It is possible that this source would be apparent at mid- to
far-IR wavelengths, but unfortunately none of the available data have
sufficient resolution to identify it. It is also notable that this
maser has an exceptionally large luminosity (7.7~L$_{\sun}$) for a
GA-\hto\/ maser. Hopefully future high resolution near to mid-IR data
may help shed light on this deeply embedded maser source.

\section{Conclusions}

We have imaged the GBT-discovered GA-\hto\/ masers in the Antennae
galaxies with exquisite angular resolution ($\sim 80$ mas), and find
two maser regions in the IAR and a third $5\farcs6$ ($\gtrsim$ 600 pc)
west of the NGC4039 nucleus.  All three masers show excellent
kinematic and spatial agreement with dense CO molecular gas. The IAR
maser regions are located in areas of high optical/near-IR extinction,
and are coincident with thermal ionized gas suggesting the presence of
several thousand O$^*$. Indeed, both of these maser regions seem to
pinpoint extremely young sites of deeply embedded super star cluster
formation. These results highlight the promise of using GA-\hto\/
masers to precisely locate the earliest phases of extragalactic
cluster formation. The current isotropic luminosity sensitivity limit
of 0.6 L$_{\sun}$, is still brighter than most Galactic \hto\/ masers,
suggesting we may only be seeing the bright tip of the maser
distribution. With its new broad bandwidth correlator, the Expanded
Very Large Array will be an ideal instrument for future GA-\hto\/
maser studies (the 1.3~cm data presented here could be taken in 1/3
the time, while achieving $10\times$ better continuum sensitivity).

%\hspace{0.5cm}

\acknowledgements This research used archival {\em Spitzer} and {\em
  Hubble} Space Telescope data, operated by the Jet Propulsion
Laboratory and Space Telescope Science Institute, respectively under
NASA contracts. We thank B. Brandl, C. Wilson, D. Iono, and
G. Petitpas for providing their published data in digital format.
K. E. J. acknowledges support from NSF through CAREER award 0548103
and the David and Lucile Packard Foundation through a Packard
Fellowship.

%\clearpage

\clearpage

\end{document}